\documentclass[conference]{IEEEtran} 

\usepackage{amssymb}

\usepackage{lineno}
\usepackage{epsfig}
\usepackage[utf8x]{inputenc}
\usepackage{times}
\usepackage{graphicx}
\usepackage[figuresright]{rotating}
\usepackage{float}
\usepackage{amssymb}
\usepackage{amsmath}
\usepackage{url}
\usepackage{hyperref}
\usepackage{soul}
\usepackage{wrapfig}
\floatname{algorithm}{Algorithm}
\usepackage[usenames]{color}
\usepackage{flushend}
\usepackage[nolist]{acronym}
\usepackage{multirow}
\usepackage{mathtools}
\DeclarePairedDelimiter{\ceil}{\lceil}{\rceil}
\usepackage{colortbl}
\makeatletter
\newcommand{\thickhline}{%
    \noalign {\ifnum 0=`}\fi \hrule height 1.5pt
    \futurelet \reserved@a \@xhline
}
\newcolumntype{"}{@{\hskip\tabcolsep\vrule width 1pt\hskip\tabcolsep}}
\makeatother
\newcolumntype{?}{!{\vrule width 1.5pt}}
\usepackage[table]{xcolor} 
\usepackage{etoolbox}

\usepackage{booktabs}
\usepackage[sort,compress]{cite}

\begin{document}
\newcommand{\squeezeupppp}{\vspace{-8 mm}}
\newcommand{\squeezeuppp}{\vspace{-6 mm}}
\newcommand{\squeezeupp}{\vspace{-5 mm}}
\newcommand{\squeezeup}{\vspace{-3 mm}}
\newcommand{\squeezeu}{\vspace{-2 mm}}
\newcommand{\squeeze}{\vspace{-1 mm}}
\newcommand{\squeez}{\vspace{-.5 mm}}

\begin{acronym}
\end{acronym}



\title{Towards Higher Spectral Efficiency: Rate-2 Full-Diversity Complex Space-Time Block Codes}




\author{\IEEEauthorblockN{Anu Jagannath, Jithin Jagannath, Andrew Drozd}

\IEEEauthorblockA{Marconi-Rosenblatt AI/ML Innovation Laboratory, ANDRO Computational Solutions, LLC, Rome, NY, 13440\\
E-mail: \{ajagannath, jjagannath, adrozd\}@androcs.com 
}

}

\maketitle

\begin{abstract}
The upcoming 5G (5\textsuperscript{th} Generation) networks demand high-speed and high spectral-efficiency communications to keep up with the proliferating traffic demands. To this end, Massive multiple-input multiple-output (MIMO) techniques have gained significant traction owing to its ability to achieve these without increasing bandwidth or density of base stations. The preexisting space-time block code (STBC) designs cannot achieve a rate of more than 1 for more than two transmit antennas while preserving the orthogonality and full diversity conditions.

In this paper, we present \textit{Jagannath codes} - a novel complex modulation STBC, that achieves a very high rate of 2 for three and four transmit antennas. The presented designs achieve full diversity and overcome the previously achieved rates with the three and four antenna MIMO systems. We present a detailed account of the code construction of the proposed designs, orthogonality and full diversity analysis, transceiver model and conditional maximum likelihood (ML) decoding. In an effort to showcase the improvement achieved with the presented designs, we compare the rates and delays of some of the known STBCs with the proposed designs. The effective spectral efficiency and coding gain of the presented designs are compared to the Asymmetric Coordinate Interleaved design (ACIOD) and Jafarkhani code. We presented an effective spectral efficiency improvement by a factor of $2$ with the proposed Jagannath codes. Owing to the full diversity of the presented designs, we demonstrate significant coding gains ($~6$ dB and $12$ dB) with the proposed designs.
\end{abstract}


\section{Introduction and Background}\label{sec:intro}
Massive multiple-input multiple-output (MIMO) has received significant attention in recent years as a key enabling technology for the 5G mobile communication systems \cite{5gbook}. Massive MIMO systems attain higher transmission rates owing to the large number of antennas being used at the base station (BS). Space-time block codes (STBC) is a well researched topic in this regard for its ability to achieve higher transmission rates by exploiting transmit antenna diversity \cite{Alamouti,orthomaxrate,Su2003,ostbcnon,ZafarRajan,3tx,frlr,golden,rate2,JafarkhaniQSTBC,WSu2004,WSu2008}. In 1998, Alamouti proposed a simple transmit diversity scheme \cite{Alamouti} which exploited two transmit antennas. The Alamouti scheme was consequently adopted in the Third Generation (3G) Mobile standard. 

STBC enables attaining higher levels of spectral efficiencies for a fixed bandwidth and error-rate. In \cite{Telatar99,Foschini1998} it has been shown that STBC can achieve phenomenal increase in capacity in contrast to single transmit/receive antenna systems. To be in pace with the rapidly growing traffic \cite{EricssonMobility2018}, an imperative design goal of 5G technologies is to improve area throughput (bits/s/km$^2$) which is directly related to the bandwidth, base station density and spectral efficiency. Among these, improving spectral efficiency without increasing the bandwidth or base station density is attainable with superior MIMO techniques. The increase in achievable spectral efficiency for massive MIMO system and the effect of number of BS transmit antennas was studied in \cite{SE2017}. The key to improving the spectral efficiency for a MIMO system is by increasing the coding rate of the STBC.  
In \cite{stbcfullrate2007}, a full-rate full-diversity $2\times2$ STBC was proposed whose detection complexity scaled quadratically with the cardinality of the signal constellation. A full-rate linear receiver (FRLR) $2\times2$ STBC was introduced in \cite{frlr} which demonstrates satisfactory performance only for binary phase shift keying (BPSK) and Quadrature Amplitude Modulation-4 (QAM) constellations. Higher order modulation support is essential for attaining higher spectral efficiency. A full rate $2\times2$ STBC referred to as Golden code was proposed in \cite{golden} but with a decoding complexity of $\mathcal{O}(Q^4)$ for a constellation cardinality of $Q$. In \cite{rate2}, a full-rate $2\times2$ STBC with low complexity conditional ML decoding was presented. Jafarkhani \cite{JafarkhaniQSTBC} proposed a $4\times4$ STBC that achieves full-rate by relaxing the orthogonality constraint. In \cite{3tx}, a $4\times3$ rate-1 STBC design was proposed. Here again, the orthogonality is compromised to achieve rate-1. Further, \cite{Tarokh1999} proposed a rate-3/4 generalized complex linear processing orthogonal design for three and four transmit antennas.

Orthogonal STBC achieves full rate and allow single complex symbol maximum-likelihood (ML) decoding for two transmit antennas. The full rate vanishes as the number of transmit antennas increase to more than two. In \cite{Tarokh1999,WSu2004}, it has been shown as per Hurwitz-Radon theorem that complex orthogonal STBC cannot possess a full rate and maximum diversity. The rate-loss with complex signal constellations while using more than two transmit antennas is the biggest drawback of orthogonal STBCs. Therefore, STBCs that achieve higher coding rates and minimal decoding complexities are preferred. It has been conjectured that for square matrix embeddable codes, the maximum achievable rate for three and four transmit antennas is $3/4$. The STBC designs presented in \cite{WSu2004,WSu2008,JafarkhaniQSTBC,rciod} achieve a rate not more than 1 for more than two transmit antennas. 
In this work, we propose -  \textit{full-diversity rate-2 orthogonal designs for three and four transmit antennas}. To the best of our knowledge, this is the first work in this domain that has successfully achieved full diversity and rate-2 for three and four transmit antenna systems.

The rest of this article is organized as follows: Section \ref{sec:stbc} gives a brief overview of STBC and associated terminologies. We present the proposed $4\times3$ and $4\times4$ Jagannath STBCs and their analysis in section \ref{sec:Jagdesigns}. Section \ref{sec:simresults} will discuss the performance comparison of the proposed designs with known STBCs. Finally, we conclude with our inferences and future works in section \ref{sec:conclude}.

\textbf{Notations:} In the presented work, we will denote vectors and matrices by lowercase and uppercase boldface letters. $(.)^H$ and $(.)^*$ denotes Hermitian transpose of a vector or matrix and complex conjugate operator. $det(.)$ is used to indicate the determinant of a matrix. The $\left|.\right|$ and $\ceil{.}$ indicate the absolute value and ceil operators. Finally, $\Re(.)$ and $\Im(.)$ denote the real and imaginary part of complex numbers.

\section{SPACE-TIME BLOCK CODES}\label{sec:stbc}
Space-time block coding refers to a channel coding technique that exploits antenna diversity. An STBC is a matrix of size $T\times N$ with real or complex symbols and its conjugates or their permutations in its entries. Here, $N$ refers to the number of transmit antennas and $T$ denotes number of epochs over which the symbols are sent from the $N$ antennas. The simplest complex orthogonal STBC proposed by Alamouti \cite{Alamouti} is
\begin{equation}
    \mathbf{C}_2 = \quad
\begin{bmatrix} 
x_1 & x_2 \\
-x_2^* & x_1^* 
\end{bmatrix}
\quad
\end{equation}
a $2\times 2$ code that transmits two symbols $x_1$ and $x_2$ over two channel uses (epochs).

\textit{Definition 1:}\textbf{ Code rate - } If a $T\times N$ STBC matrix transmits $S$ symbols over $T$ channel uses, then the code rate is defined as $S/T$ symbols per channel use (symbols/s/Hz). 
Now, it is straightforward to follow that the Alamouti code $\mathbf{C}_2$ provides a rate of $R = 2/2 = 1$.
Further, in \cite{Tirkkonen2000,Tirkkonen2001} the maximal rate of a square matrix embeddable orthogonal STBC was found to be
\begin{equation}
    R_{max} = \frac{\ceil{\log_2N + 1}}{2\ceil{\log_2N}}
\end{equation}
To achieve a rate beyond this established bound, the orthogonality will need to be sacrificed.

\textit{Definition 2:}\textbf{ Orthogonality - } A generalized complex $T\times N$ STBC matrix $\mathbf{C}$ with entries drawn from the set $\{0,\pm x_1,\cdots,\pm x_n,\cdots,\pm x_1^*,\cdots,\pm x_n^*\}$ or their product with $i=\sqrt{-1}$ is said to be orthogonal \cite{Tarokh1999} if $\mathbf{C}^H\mathbf{C} = \mathbf{D}$, where $\mathbf{D}$ is a diagonal matrix with $j^{th}$ diagonal entry 
\begin{equation}
    \mathbf{D}(j,j) = (c_1^j\left| x_1 \right| ^2 + c_2^j\left| x_2 \right| ^2 + \cdots +c_n^j\left| x_n \right| ^2)
\end{equation}
where the coefficients $\{c_1^j,c_2^j,\cdots,c_n^j\}$ are strictly positive numbers.
Similarly, a $T\times N$ generalized real orthogonal STBC $\mathbf{C}_\mathbb{R}$ would have entries drawn set of real numbers $\{0,\pm x_1,\cdots,\pm x_n\}$ and diagonal matrix $\mathbf{D}_\mathbb{R}$ with $j^{th}$ diagonal entry 
\begin{equation}
    \mathbf{D_\mathbb{R}}(j,j) = (c_1^j x_1^2 + c_2^jx_2^2 + \cdots +c_n^jx_n^2)
\end{equation}
and coefficients $\{c_1^j,c_2^j,\cdots,c_n^j\}$ are strictly positive numbers.
Considering a Rayleigh flat-fading channel $\mathbf{H}\in \mathbb{C}^{N\times N}$ with independent identically distributed (i.i.d) entries from $\mathcal{CN}(0,1)$, the received symbol matrix for a $N\times N$ MIMO transmission can be modeled as 
\begin{equation}
    \mathbf{Y} = \sqrt{\frac{\rho}{N}}\mathbf{CH} + \mathbf{N}
\end{equation}
where $\mathbf{Y}\in \mathbb{C}^{T\times N}$ is the received signal matrix, $\mathbf{C}\in \mathbb{C}^{T\times N}$ is the STBC matrix, $\mathbf{N}\in \mathbb{C}^{T\times N}$ is the additive white Gaussian noise matrix with i.i.d entries from $\mathcal{CN}(0,N_0)$. Assuming perfect channel state information (CSI) at the receiver, the ML decoding metric can be expressed as 
\begin{equation}
    \Hat{x} = \arg\underset{x}{\min}\left|\left|\mathbf{Y}-\mathbf{CH}\right|\right|_F^2 \label{eq:ML}
\end{equation}
Here, if $x$ is drawn from a constellation with cardinality $Q$, the ML decoding complexity is given by $\mathcal{O}(Q)$.

\textit{Definition 3:} \textbf{Decoding Complexity  - } The minimum number of symbols that need to be jointly decoded in minimizing the decoding metric defines the decoding complexity of a MIMO system. A decoding complexity of $\mathcal{O}(Q^k)$ implies an exhaustive search over $k$ information symbols from a signal constellation with cardinality $Q$. Here $\mathcal{O}(.)$ denotes the big omicron. ML decoding that can be expressed by the form in equation (\ref{eq:ML}) is also referred to as \textit{single-symbol decodable}.

\textit{Definition 4:} \textbf{Spectral Efficiency  - } The measure of amount of useful bits that are transmitted per channel use defines the spectral efficiency of a STBC and can be expressed as
\begin{equation}
    \eta = R\log_2Q \;\text{bits/s/Hz}
\end{equation}
An STBC with a higher coding rate will, therefore, improve the spectral efficiency of the MIMO system for a given modulation.

\textit{Definition 5:} \textbf{Coding Delay  - } The number of epochs over which the symbols of an STBC are transmitted is referred to as the coding delay. This is the same as the block length of the STBC. For a $T\times N$ STBC, the coding delay or block length is $T$. 

The $4\times4$ STBC proposed by Jafarkhani \cite{JafarkhaniQSTBC} is a Quasi-orthogonal design that builds upon the Alamouti code. Jafarkhani code achieves a rate 1 by transmitting four complex symbols over four channel uses. Let us denote the Alamouti encoding of symbols $x_1$ and $x_2$ as $\mathbf{C}_{12} = \mathbf{C}_{2}$. Now, the $4\times4$ Jafarkhani STBC corresponding to symbols $x_1$, $x_2$, $x_3$ and $x_4$ take the form,
\begin{equation}
    \mathbf{C}_{J} =
\begin{bmatrix} 
\mathbf{C}_{12} & \mathbf{C}_{34} \\
-\mathbf{C}_{34*} & \mathbf{C}_{12*} 
\end{bmatrix}
\end{equation}

The Jafarkhani design demonstrates the rate increase achieved by relaxing the orthogonality constraint. Hence, defying the $R_{max}$ bound for complex orthogonal designs. 

A rate-1 $4\times3$ non-orthogonal STBC design was proposed in \cite{3tx}. The structure again uses combinations of Alamouti structure to transmit four symbols over four epochs from three transmit antennas as,
\begin{equation}
    \mathbf{C}_{NO} =
\begin{bmatrix} 
\mathbf{C}_{12} & -\mathbf{c}_{34} \\
\mathbf{C}_{34} &  \mathbf{c}_{12} 
\end{bmatrix}
\end{equation}
where $\mathbf{c}_{12} = [x_1 \;\;x_2]^T$ and $\mathbf{c}_{34} = [x_3 \;\;x_4]^T$. It is straightforward to realize that $R=4/4 =1$ symbol/s/Hz.

Another class of STBC worth mentioning is the coordinate-interleaved orthogonal design (CIOD). CIOD is obtained by interleaving the coordinates of the symbols as proposed by \cite{rciod}. For instance, lets look at the CIOD for four transmit antennas as represented by equation (\ref{eq:ciod4}). Here, the quadrature coordinates of the symbols $\{x_1, x_2, x_3, x_4\}$ are interleaved. Notice the orthogonality of this design,
\begin{equation}
\mathbf{C}_{CIOD}^H\mathbf{C}_{CIOD}  = \begin{bmatrix}
A\mathbf{I}_2 &\mathbf{0}_2 \\
\mathbf{0}_2 & B\mathbf{I}_2
\end{bmatrix}  
\end{equation}
where $\mathbf{I}_2$ and $\mathbf{0}_2$ are the $2\times2$ identity and zero matrices. Here, $A=\Re(x_0)^2 + \Re(x_1)^2 + \Im(x_2)^2 + \Im(x_3)^2 $ and $B=\Re(x_2)^2 + \Re(x_3)^2 + \Im(x_0)^2 + \Im(x_1)^2$. It is intuitive that two different columns are orthogonal to each other while the standard dot product of different columns are different. It is also observable that no cross terms of the form $\{\Re(x_k)\Re(x_l)\}$ exist in $A, B$ implying single-symbol decodability. The authors of \cite{rciod} also proposed a $4\times3$ STBC design that derived from the $4\times4$ design in equation (\ref{eq:ciod4}) by deleting the fourth column. The design hence obtained is referred to as the Asymmetric CIOD - ACIOD.
\begin{figure*}
\begin{equation}
\mathbf{C}_{CIOD} =
\begin{bmatrix} 
\Re(x_1)+ j\Im(x_3) & \Re(x_2)+ j\Im(x_4) &0 &0\\
-\Re(x_2)+ j\Im(x_4) & \Re(x_1) - j\Im(x_3) &0 &0\\
0 &0 &\Re(x_3) + j\Im(x_1) &\Re(x_4) + j\Im(x_2)\\
0 &0 &-\Re(x_4) + j\Im(x_2) & \Re(x_3) - j\Im(x_1) 
\end{bmatrix} \label{eq:ciod4}
\end{equation}
\end{figure*}
\section{PROPOSED ORTHOGONAL DESIGNS}\label{sec:Jagdesigns}
\subsection{\textbf{Jagannath} $\mathbf{4\times3}$ \textbf{STBC}}
The proposed rate-2 orthogonal design for three transmit antennas is $\mathbf{C}^{P3}$ as in equation (\ref{eq:CP3}). 
\begin{figure*}
\begin{equation}
\mathbf{C}_{P3} = \quad
\begin{bmatrix} 
0 & x_1\sin{\alpha_1}-x_2^*\cos{\alpha_1} & x_3\sin{\alpha_2}-x_4^*\cos{\alpha_2} \\
0 & -x_3^*\sin{\alpha_2}+x_4\cos{\alpha_2} & x_1^*\sin{\alpha_1}-x_2\cos{\alpha_1} \\
x_5\sin{\alpha_1}-x_6^*\cos{\alpha_1} & x_7\sin{\alpha_1}-x_8^*\cos{\alpha_1} & 0\\
-x_7^*\sin{\alpha_1}+x_8\cos{\alpha_1} & x_5^*\sin{\alpha_1}-x_6\cos{\alpha_1} & 0 
\end{bmatrix} \label{eq:CP3}
\quad 
\end{equation}
\squeezeup
\end{figure*}
Here, we encode eight symbols for transmission from three transmit antennas over four channel uses resulting in a rate-2 transmission. 
For ease of reference, let us denote the two symbol encoding as $J_{x_i,x_{i+1}}^{t} = x_i\sin{\alpha_t}-x_{i+1}^*\cos{\alpha_t} $. The orthogonality of the proposed design can be verified as 
\begin{equation}
    \mathbf{C}_{P3}^H \mathbf{C}_{P3} = \begin{bmatrix}
D &0 &0 \\
0 &F &0 \\
0 &0 & C
\end{bmatrix} 
\end{equation}
where $C=\left| J_{x_1,x_2}^1\right| ^2 + \left| J_{x_3,x_4}^2\right| ^2$, $D=\left| J_{x_5,x_6}^1\right| ^2 + \left| J_{x_7,x_8}^2\right| ^2$ and $F=C+D$. The columns of $\mathbf{C}_{P3}$ are orthogonal to each other with the standard dot product of different columns are different. 
Lets take a closer look at the received signal model and the decoding. Consider a $3\times3$ MIMO system with channel matrix $\mathbf{H}_3$ with i.i.d channel coefficients from $\mathcal{CN}(0,N_0)$.\;
 Each row of the channel matrix corresponds to the channel vector between the three transmit antennas and the receive antenna at the receiver. For ease of convenience, we will denote each row as $\mathbf{h}_{r} = [h_{0r}, h_{1r}, h_{2r}]$, where $r=\{0, 1,2\}$ represents the row and indexes the receive antenna. The received signal at the $i^{th}$ receive antenna at the four epochs is as in equation \ref{eq:zi}. 
\begin{equation}
\quad
\begin{bmatrix} 
z_i^1 \\
z_i^2 \\
z_i^3\\
z_i^4
\end{bmatrix}\quad = \sqrt{\frac{\rho}{3}}\mathbf{C}_{P3}\mathbf{h}_{r} + \quad\begin{bmatrix} 
n_i^1 \\
n_i^2 \\
n_i^3\\
n_i^4
\end{bmatrix}\quad\label{eq:zi}
\end{equation}

This can be rewritten in the equivalent virtual channel matrix (EVCM) form as

\begin{equation}
\begin{bmatrix} 
z_i^1\\
z_i^{2*}
\end{bmatrix}
  =
\sqrt{\frac{\rho}{3}}\begin{bmatrix} 
h_{1i} & h_{2i}\\
h_{2i}^* & -h_{1i}^*
\end{bmatrix}\begin{bmatrix} 
x_1\sin{\alpha_1}-x_2^*\cos{\alpha_1}\\
x_3\sin{\alpha_2}-x_4^*\cos{\alpha_2}
\end{bmatrix} + \begin{bmatrix} 
n_i^1\\
n_i^{2*}
\end{bmatrix}
\end{equation}

\begin{equation}
\begin{bmatrix} 
z_i^3\\
z_i^{4*}
\end{bmatrix}
=
\sqrt{\frac{\rho}{3}}\begin{bmatrix} 
h_{0i} & h_{1i}\\
h_{1i}^* & -h_{0i}^*
\end{bmatrix}\begin{bmatrix} 
x_5\sin{\alpha_1}-x_6^*\cos{\alpha_1}\\
x_7\sin{\alpha_2}-x_8^*\cos{\alpha_2}
\end{bmatrix} + \begin{bmatrix} 
n_i^3\\
n_i^{4*}
\end{bmatrix}
\end{equation}

Assuming perfect CSI, the channel equalization would result in

\begin{align}
\begin{bmatrix} 
q_i^1\\
q_i^2
\end{bmatrix}
&=
\sqrt{\frac{\rho}{3}}\begin{bmatrix} 
h_{1i}^* & h_{2i}\\
h_{2i}^* & -h_{1i}
\end{bmatrix}\begin{bmatrix} 
z_i^1\\
z_i^{2*}
\end{bmatrix}\\ \nonumber
&=\sqrt{\frac{\rho}{3}}(\left| h_{1i}\right|^2 + \left| h_{2i}\right|^2)\begin{bmatrix} 
x_1\sin{\alpha_1}-x_2^*\cos{\alpha_1}\\
x_3\sin{\alpha_2}-x_4^*\cos{\alpha_2}
\end{bmatrix} + \begin{bmatrix} 
g_i^1\\
g_i^{2*}
\end{bmatrix}
\\
\begin{bmatrix} 
q_i^3\\
q_i^4
\end{bmatrix}
  &=
\sqrt{\frac{\rho}{3}}\begin{bmatrix} 
h_{0i}^* & h_{1i}\\
h_{1i}^* & -h_{0i}
\end{bmatrix}\begin{bmatrix} 
z_i^3\\
z_i^{4*}
\end{bmatrix}\\ \nonumber
&=\sqrt{\frac{\rho}{3}}(\left| h_{0i}\right|^2 + \left| h_{1i}\right|^2)\begin{bmatrix} 
x_5\sin{\alpha_1}-x_6^*\cos{\alpha_1}\\
x_7\sin{\alpha_2}-x_8^*\cos{\alpha_2}
\end{bmatrix} + \begin{bmatrix} 
g_i^3\\
g_i^{4*}
\end{bmatrix}
\end{align}

Now, the sufficient statistic to jointly estimate the symbols $x_1$ and $x_2$ is
\begin{equation}
\beta^1 =\frac{1}{3}\sum_{l=0}^2 q_l^1.
\end{equation}
Likewise, the sufficient statistics to estimate the symbol pairs $\{x_3,x_4\}$, $\{x_5,x_6\}$, and $\{x_7,x_8\}$ are
\begin{equation}
    \beta^2 =\frac{1}{3}\sum_{l=0}^2  q_l^2, \; \beta^3 =\frac{1}{3}\sum_{l=0}^2 q_l^3, \; \text{and}\;\beta^4 =\frac{1}{3}\sum_{l=0}^2 q_l^4 
\end{equation}

   
To allow conditional ML decoding from the sufficient statistic, we will construct intermediate signals corresponding to each as follows,
\begin{equation}
\Tilde{\beta^i} = \beta^i - \sqrt{\frac{\rho}{27}}\Psi_m[-x_{2i}^*\cos{\alpha_j}]
\end{equation}
where  $x_{2i}$ is one of the $Q$ constellation points, $i=\{1, 2, 3, 4\}$ denotes the epoch, $m=\{1, 2\}$ and $j=\{1, 2\}$ takes values such that

\begin{equation}
m = \left\{ \,
\begin{IEEEeqnarraybox}[][c]{l?s}
\IEEEstrut
1 & if $i=\{1, 2\}$, \\
2 & if $i=\{3, 4\}$.
\IEEEstrut
\end{IEEEeqnarraybox}
\right.
\label{eq:m}
\end{equation}

\begin{equation}
j = \left\{ \,
\begin{IEEEeqnarraybox}[][c]{l?s}
\IEEEstrut
1 & if $i=\{1, 3\}$, \\
2 & if $i=\{2, 4\}$.
\IEEEstrut
\end{IEEEeqnarraybox}
\right.
\label{eq:mm}
\end{equation}
Here $\Psi_1 = \sum_{p=0}^2(\left| h_{1p}\right|^2 + \left| h_{2p}\right|^2)$ and $\Psi_2 = \sum_{p=0}^2(\left| h_{0p}\right|^2 + \left| h_{1p}\right|^2)$ respectively. The values for $\alpha_1$ and $\alpha_2$ are chosen as in \cite{rate2} to maximize the coding gain. The ML estimate of the symbols $x_1, x_3, x_5, x_7$ conditional on $x_2, x_4, x_6, x_8$ respectively denoted by $x_{2i-1|2i}$ are obtained by feeding the intermediate signals to a threshold detector. For each of the $Q$ constellation points, the conditional ML estimate that minimizes the following cost function yields the correct symbol pair.

\begin{equation}
    \tau^i = \left|\beta^i - \sqrt{\frac{\rho}{12}}\Psi_m \left[ x_{2i-1|2i}\sin{\alpha_j}-x_{2i}^*\cos{\alpha_j} \right]\right|^2
\end{equation}

\subsubsection{\underline{Full Diversity Analysis}}
Lets suppose that the two distinct $4\times3$ codeword matrices be $\mathbf{X}$ and $\mathbf{U}$ such that $\mathbf{X}$ is constructed from entries $\{J_{x_1,x_2}^1, J_{x_3,x_4}^2,J_{x_5,x_6}^1, J_{x_7,x_8}^2\}$ and $\mathbf{U}$ from $\{J_{u_1,u_2}^1, J_{u_3,u_4}^2,J_{u_5,u_6}^1, J_{u_7,u_8}^2\}$. The difference matrix $\mathbf{X-U})_{P3}$ must be full rank for any two different codewords \cite{frlr,TarokhRank}. The difference matrix can be obtained as
\begin{equation}
(\mathbf{X-U})_{P3} = 
\begin{bmatrix} 
0 & J_{d_1,d_2}^1 & J_{d_3,d_4}^2 \\
0 & -J_{d_3,d_4}^{2*} & J_{d_1,d_2}^{1*} \\
J_{d_5,d_6}^1 & J_{d_7,d_8}^2 & 0\\
-J_{d_7,d_8}^{2*} & J_{d_5,d_6}^{1*} & 0 
\end{bmatrix}. \label{eq:diff3}
\end{equation}
Now, we have
\begin{equation} \label{eq:diverse3}
\begin{split}
 det\{(\mathbf{X-U})_{P3}^H(\mathbf{X-U})_{P3}\} =(\left| J_{d_5,d_6}^1\right| ^2 + \left| J_{d_7,d_8}^2\right|^2)\times\\(\left| J_{d_1,d_2}^1\right| ^2 + \left| J_{d_3,d_4}^2\right| ^2 + \left| J_{d_5,d_6}^1\right| ^2 + \left| J_{d_7,d_8}^2\right| ^2)\times&\\(\left| J_{d_1,d_2}^1\right| ^2 + \left| J_{d_3,d_4}^2\right| ^2).
\end{split}
\end{equation}
where $J_{d_i,d_{i+1}}^j = (x_1-u_1)\sin{\alpha_j} - (x_2-u_2)^*\cos{\alpha_j}$. It can be easily verified that the three terms of equation (\ref{eq:diverse3}) are positive scalars. Consequently, the proposed $4\times3$ STBC achieves full-rank and hence full diversity.
\subsection{\textbf{Jagannath} $\mathbf{4\times 4}$\;\textbf{STBC}} 
The proposed rate-2 STBC for four transmit antennas is
$\mathbf{C}_{P4}$ as in equation (\ref{eq:CP4}). The orthogonality of the proposed rate-2 STBC can be verified by 
\begin{equation}
    \mathbf{C}_{P4}^H \mathbf{C}_{P4} = \begin{bmatrix}
C\mathbf{I}_2 &\mathbf{0}_2 \\
\mathbf{0}_2 & D\mathbf{I}_2
\end{bmatrix} 
\end{equation}
Here, the columns are orthogonal to each other but the dot product of different columns are different.
The channel matrix of the $4\times 4$ MIMO system can be denoted as $\mathbf{H}^{4}$, whose row $\mathbf{h}_{r} = [h_{0r}, h_{1r}, h_{2r}, h_{3r}]$, $r=\{0, 1,2, 3\}$.\;
Now, the received signal at the $i^{th}$ antenna during the four epochs can be represented as
\begin{figure*}
\begin{equation}
\mathbf{C}_{P4} = \quad
\begin{bmatrix} 
x_1\sin{\alpha_1}-x_2^*\cos{\alpha_1} & x_3\sin{\alpha_2}-x_4^*\cos{\alpha_2} &0 &0\\
-x_3^*\sin{\alpha_2}+x_4\cos{\alpha_2} & x_1^*\sin{\alpha_1}-x_2\cos{\alpha_1} &0 &0\\
0 &0 &x_5\sin{\alpha_1}-x_6^*\cos{\alpha_1} & x_7\sin{\alpha_1}-x_8^*\cos{\alpha_1}\\
0 &0 &-x_7^*\sin{\alpha_1}+x_8\cos{\alpha_1} & x_5^*\sin{\alpha_1}-x_6\cos{\alpha_1} 
\end{bmatrix} \label{eq:CP4}
\quad 
\end{equation}
\end{figure*}
\begin{equation}
\quad
\begin{bmatrix} 
z_i^1 \\
z_i^2 \\
z_i^3\\
z_i^4
\end{bmatrix}\quad = \sqrt{\frac{\rho}{4}}\mathbf{C}_{P4}\mathbf{h}_{r} + \quad\begin{bmatrix} 
n_i^1 \\
n_i^2 \\
n_i^3\\
n_i^4
\end{bmatrix}\quad\label{eq:zi4}
\end{equation}
In a similar  manner to the $4\times3$ STBC, we will rewrite the equation (\ref{eq:zi4}) to the EVCM form as in equation (\ref{eq:evcm4}).
\begin{figure*}
\begin{equation}
\quad
\begin{bmatrix} 
z_i^1\\
z_i^{2*}\\
z_i^3\\
z_i^{4*}
\end{bmatrix}
= 
\sqrt{\frac{\rho}{4}}\begin{bmatrix} 
h_{0i} & h_{1i} &0 &0\\
h_{1i}^* & -h_{0i}^* &0 &0\\
0 &0 &h_{2i} & h_{3i}\\
0 &0 &h_{3i}^* & -h_{2i}^*
\end{bmatrix}\begin{bmatrix} 
x_1\sin{\alpha_1}-x_2^*\cos{\alpha_1}\\
x_3\sin{\alpha_2}-x_4^*\cos{\alpha_2}\\
x_5\sin{\alpha_1}-x_6^*\cos{\alpha_1}\\
x_7\sin{\alpha_1}-x_8^*\cos{\alpha_1}
\end{bmatrix} + \begin{bmatrix} 
n_i^1\\
n_i^{2*}\\
n_i^3\\
n_i^{4*}
\end{bmatrix}\label{eq:evcm4}
\quad 
\end{equation}
\squeezeupp
\end{figure*}
The channel equalization would yield the expression in equation (\ref{eq:equa4}). The sufficient statistics and the intermediate symbol representation to decode the symbol pairs can be obtained in a similar manner as
\begin{equation}
\beta^i =\frac{1}{4}\sum_{l=0}^3 q_l^i.
\end{equation}

\begin{equation}
    \Tilde{\beta^i} =\beta^i - \sqrt{\frac{\rho}{64}}\Psi_m[-x_{2i}^*\cos{\alpha_j}]
\end{equation}

\begin{equation}
    \tau^i = \left|\beta^i - \sqrt{\frac{\rho}{64}}\Psi_m\left[x_{2i-1|2i}\sin{\alpha_j}-x_{2i}^*\cos{\alpha_j}\right]\right|^2 \label{eq:cost4}
\end{equation}
For each of the $Q$ constellation points, the conditional ML estimate ($x_{2i-1|2i}$) that minimizes the cost function (\ref{eq:cost4}) yields the correct symbol pair. Here, $i, m , j$ mean the same notations as in the $4\times3$ STBC case while 
\begin{equation}
\Psi_1 = \sum_{p=0}^3(\left| h_{0p}\right|^2 + \left| h_{1p}\right|^2), \;\Psi_2 = \sum_{p=0}^3(\left| h_{2p}\right|^2 + \left| h_{3p}\right|^2).
\end{equation}
\begin{figure*}
\begin{equation}
\begin{bmatrix} 
q_i^1\\
q_i^{2}\\
q_i^3\\
q_i^{4}
\end{bmatrix}
= 
\sqrt{\frac{\rho}{4}}\begin{bmatrix} 
(\left| h_{0i}\right|^2 + \left| h_{1i}\right|^2) &0 &0 &0\\
0 &(\left| h_{0i}\right|^2 + \left| h_{1i}\right|^2) &0 &0\\
0 &0 &(\left| h_{2i}\right|^2 + \left| h_{3i}\right|^2) &0\\
0 &0 &0 &(\left| h_{2i}\right|^2 + \left| h_{3i}\right|^2)
\end{bmatrix}\begin{bmatrix} 
x_1\sin{\alpha_1}-x_2^*\cos{\alpha_1}\\
x_3\sin{\alpha_2}-x_4^*\cos{\alpha_2}\\
x_5\sin{\alpha_1}-x_6^*\cos{\alpha_1}\\
x_7\sin{\alpha_1}-x_8^*\cos{\alpha_1}
\end{bmatrix} + \begin{bmatrix} 
g_i^1\\
g_i^{2}\\
g_i^3\\
g_i^{4}
\end{bmatrix}\label{eq:equa4}
\end{equation}

\end{figure*}
The conditional ML decoding procedure presented for both the proposed designs presents very low decoding complexity of $\mathcal{O}(Q)$.
A noticeable tradeoff of the proposed designs is the unequal energy on the antennas due to the transmission of zeros in the codeword. The energy can be normalized by multiplying the $4\times4$ STBC with a normalized Hadarmard matrix of order 4 prior to transmission and performing the reverse operation by multiplying by the transpose of Hadamard matrix. Similarly, the $4\times 3$ STBC can be efficiently precoded to minimize the peak to average power ratio. The appropriate precoding for both the proposed $4\times 3$ and $4\times4$ designs will be the subject of our future research.  
\subsubsection{\underline{Full Diversity Analysis}}
The full diversity characteristics of the proposed $4\times4$ STBC will be analyzed in this section. Let the two distinct $4\times4$ codeword matrices be $\mathbf{X}$ and $\mathbf{U}$ each formed of entries $\{J_{x_1,x_2}^1, J_{x_3,x_4}^2,J_{x_5,x_6}^1, J_{x_7,x_8}^2\}$ and $\{J_{u_1,u_2}^1, J_{u_3,u_4}^2,J_{u_5,u_6}^1, J_{u_7,u_8}^2\}$ respectively. Let the difference matrix $(\mathbf{X-U})_{P4}$ contain the elements $\{J_{d_1,d_2}^1, J_{d_3,d_4}^2,J_{d_5,d_6}^1, J_{d_7,d_8}^2\}$. Here, we can express the full diversity criterion as,
\begin{equation} \label{eq:diverse4}
\begin{split}
 &\left| det\{(\mathbf{X-U})_{P4}\}\right| ^2 =\\ 
&=\left|(\left| J_{d_1,d_2}^1 \right|^2 + \left| J_{d_3,d_4}^2\right|^2)\right|^2 \left|(\left| J_{d_5,d_6}^1 \right|^2 + \left| J_{d_7,d_8}^2\right|^2)\right|^2 
\end{split}
\end{equation}
Clearly, equation (\ref{eq:diverse4}) is a positive scalar. Hence, the full diversity of the proposed $4\times4$ STBC is lucidly stated.

In Table \ref{table:rate&delay}, we compare the rates and delays of some of the known STBCs with the proposed designs.
\begin{table}[h!]
\small
    \centering
    \caption{Comparison of Rate and Delay of known STBCs with the Proposed designs\label{table:rate&delay}}{%
    \begin{tabular}{lccc} 
    \toprule
      \textbf{Design} &\textbf{TX antennas}   & \textbf{Rate}  & \textbf{Delay}\\
      \midrule
      Jagannath $4\times3$  &3 & 2 &  4   \\ 
      Jagannath $4\times4$ &4 & 2 &  4   \\ 
      ACIOD\cite{rciod}  &3 & 1 &  4   \\
      CIOD\cite{rciod}  &4 & 1 &  4   \\
      Jafarkhani\cite{JafarkhaniQSTBC} &4 &1 &4\\
      Ozbek.et.al\cite{3tx} &3 &1 &4\\
      Tarokh et. al \cite{Tarokh1999} &3 &3/4 &4\\
      Tarokh et. al \cite{Tarokh1999} &4 &3/4 &4\\
      Grover et. al\cite{SuTx4} &4 &1 &8\\ 
      \bottomrule
    \end{tabular}}{}
    
\end{table}
It is noticeable that the proposed designs offer very high rate of 2 in comparison to the known STBCs without exceeding the minimum delay presented by the compared schemes.
%






\section{SIMULATION RESULTS}\label{sec:simresults}
\begin{figure}[h!]
\centering
\epsfig{file=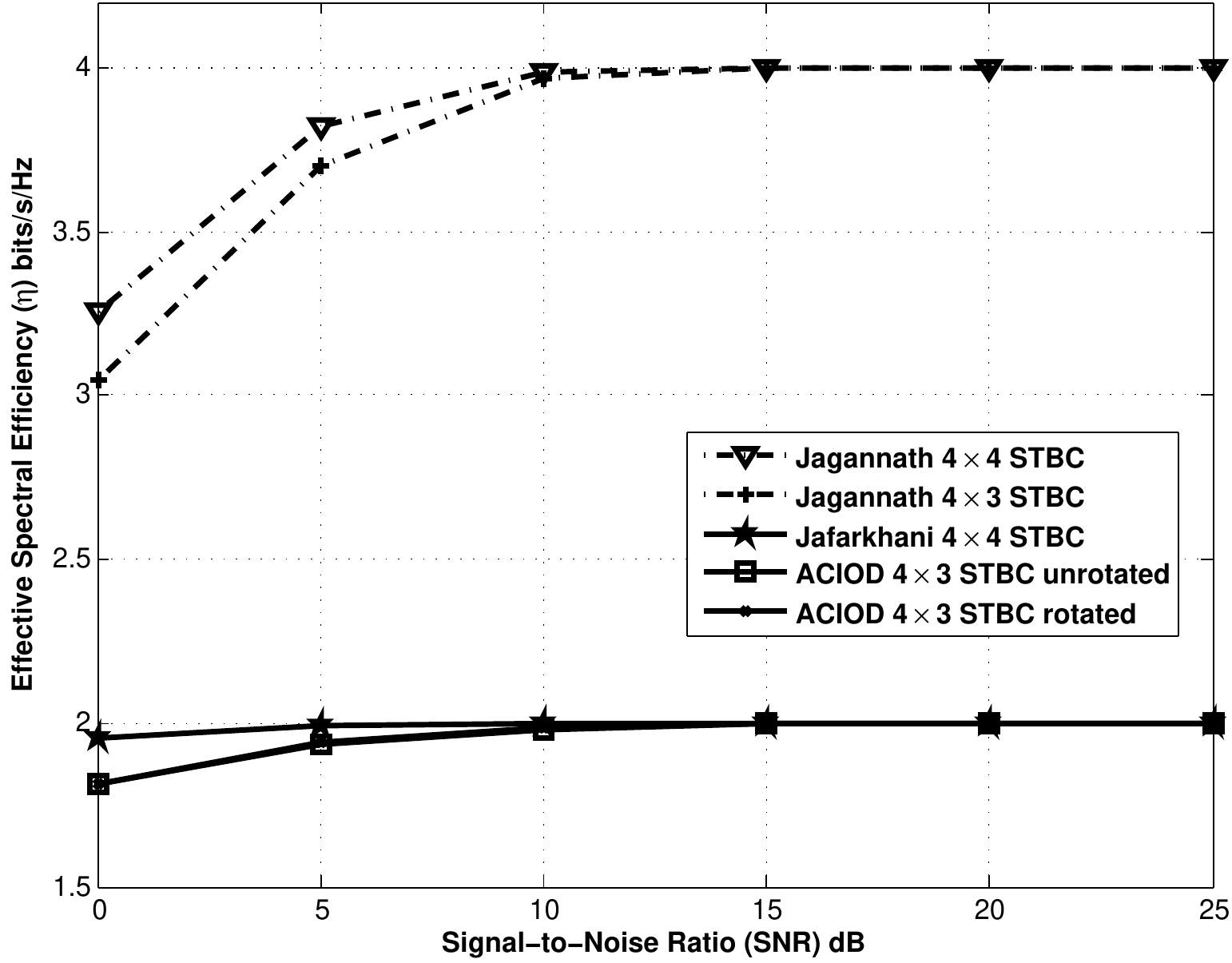, width=3.1 in,}
\caption{Spectral efficiency performance of the proposed designs}\label{fig:seff_qam4}
\end{figure}

In this section, we present the simulation results of the proposed $4\times3$ and $4\times4$ Jagannath codes with some known designs. We will use the effective spectral efficiency defined as,
\begin{equation}
\eta = [1-SER]R\log_2{Q}    
\end{equation}
and Signal-to-Noise ratio (SNR)/Coding gain as the key performance metrics to benchmark the proposed designs. Here, $SER$ denotes the symbol error rate. For three transmit antennas, we compare the proposed $4\times3$ STBC with the ACIOD design for three transmit antennas. While the proposed $4\times4$ design is compared to the Jafarkhani $4\times4$ STBC. The simulations are performed with flat-fading Rayleigh channel in additive white Gaussian noise with i.i.d as mentioned in section \ref{sec:stbc}. Each data point on the curve is an average over 10,000 repetitions. Figure \ref{fig:seff_qam4} compares the effective spectral efficiency of the proposed designs with that of ACIOD and Jafarkhani using QAM-4 modulation scheme.  With QAM-4 modulation, the maximum achievable spectral efficiency for the proposed designs is $4$ bits/s/Hz while that of ACIOD and Jafarkhani are $2$ bits/s/Hz. All designs achieve their maximum achievable spectral efficiency at an SNR of $10$ dB and above. The spectral efficiency gain by a factor of $2$ achieved with the proposed designs is noticeable in Fig.\ref{fig:seff_qam4}. 

\begin{figure}[h!]
\centering
\epsfig{file=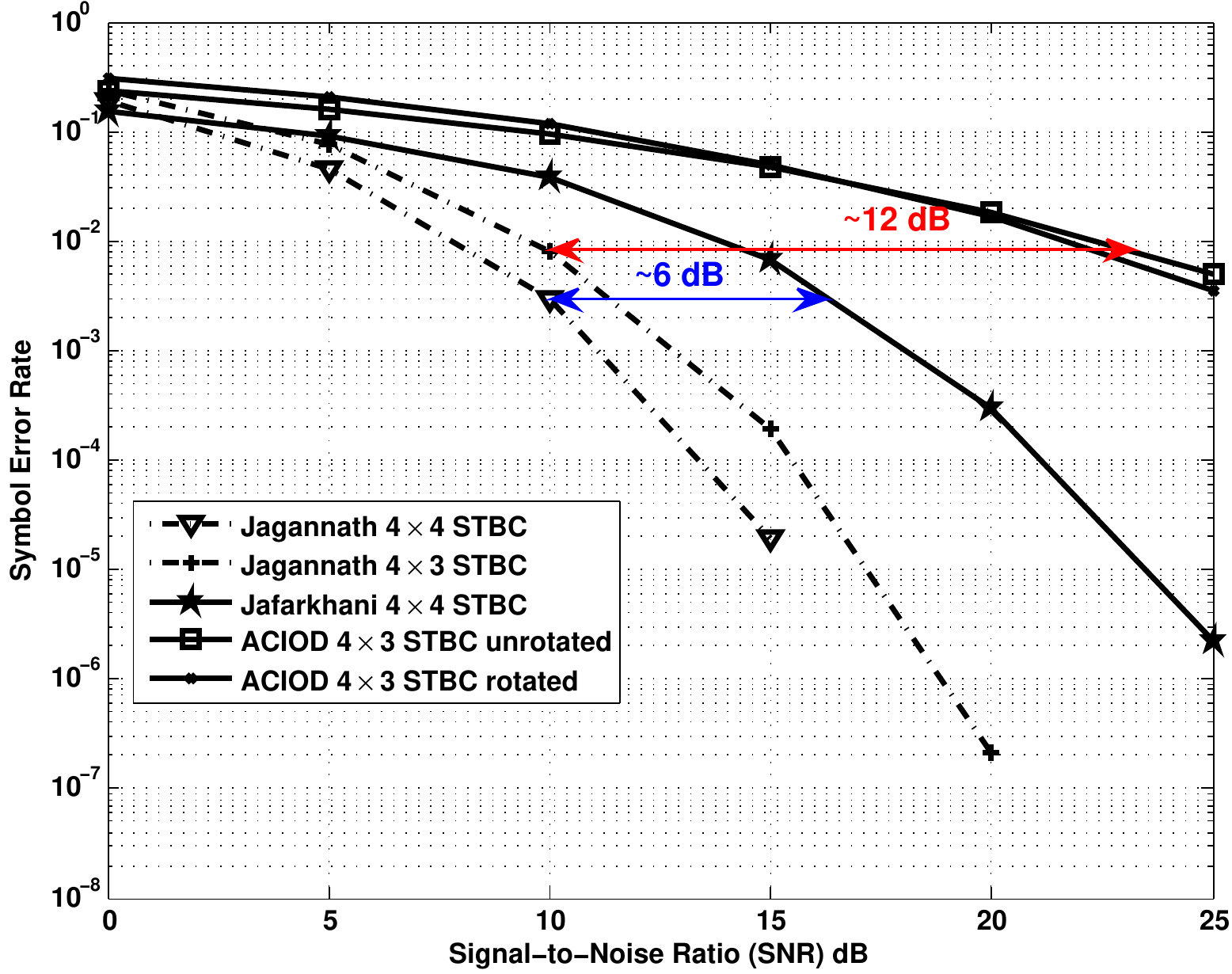, width=3.1 in,}
\caption{SNR Gain of Proposed designs at 4 bits/s/Hz spectral efficiency}\label{fig:snrgain}
\squeezeupp
\end{figure}
Figure \ref{fig:snrgain} benchmarks the SER performance of the proposed designs at a fixed spectral efficiency ($4$ bits/s/Hz) with that of ACIOD and Jafarkhani. The ACIOD and Jafarkhani use QAM-16 to attain a spectral efficiency of $4$ bits/s/Hz. We compare the rotated and unrotated versions of ACIOD since the authors of \cite{rciod} has shown that their design achieves full diversity only when the QAM constellation is rotated at an angle $31.7175^{\circ}$. Even though all designs start off at a comparable SER, the performance of the proposed Jagannath codes significantly improves with increasing SNR owing to their full diversity property. The substantial coding gain achieved with the proposed designs is observable at approximately $6$ dB and $12$ dB for the $4\times4$ and $4\times3$ configurations respectively.  This performance gain can be attributed to the full diversity property of the proposed designs while the Jafarkhani design only provides partial diversity. The unrotated version of ACIOD cannot achieve full diversity and hence performs poorly. Comparing the rotated and unrotated versions of ACIOD with each other, the benefit from rotating the constellation can be faintly noted at higher SNR values.  
\section{CONCLUSION AND FUTURE WORKS}\label{sec:conclude}
This work presented very high rate (2) Jagannath codes that achieves full diversity pertaining to three and four transmit antennas. We have detailed the STBC construction, orthogonality and diversity analysis, transceiver model, and decoding of the proposed designs. The conditional ML decoding presented a low decoding complexity of $\mathcal{O}(Q)$. Such a design with very high rate and low complexity decoding are generally preferred in practical applications. To the best of our knowledge, this is the first work in this realm that presented rate-2 designs for three and four transmit antenna systems. The previously known designs \cite{rciod,Tarokh1999,3tx,SuTx4,JafarkhaniQSTBC} for three and four transmit antennas could not support a rate more than 1. We have demonstrated the increased spectral efficiency and coding gains achieved with the proposed designs in comparison to the Jafarkhani \cite{JafarkhaniQSTBC} and ACIOD \cite{rciod}. The future work will entail appropriate precoding strategies for the proposed designs and extending the designs to higher antenna configurations.

\small
\bibliographystyle{ieeetr}
\bibliography{IDAD}

\begin{thebibliography}{10}

\bibitem{5gbook}
W.~Xiang, K.~Zheng, and X.~S. Shen, {\em 5G Mobile Communications}.
\newblock Springer Publishing Company, Incorporated, 1st~ed., 2016.

\bibitem{Alamouti}
S.~M. {Alamouti}, ``A simple transmit diversity technique for wireless
  communications,'' {\em IEEE Journal on Selected Areas in Communications},
  vol.~16, pp.~1451--1458, Oct 1998.

\bibitem{orthomaxrate}
{X.B. Liang}, ``Orthogonal designs with maximal rates,'' {\em IEEE Transactions
  on Information Theory}, vol.~49, pp.~2468--2503, Oct 2003.

\bibitem{Su2003}
W.~Su and X.~G. Xia, ``On space-time block codes from complex orthogonal
  designs,'' {\em Wireless Personal Communications}, vol.~25, pp.~1--26, Apr
  2003.

\bibitem{ostbcnon}
{X. B. Liang} and {X. G. Xia}, ``On the nonexistence of rate-one generalized
  complex orthogonal designs,'' {\em IEEE Transactions on Information Theory},
  vol.~49, pp.~2984--2988, Nov 2003.

\bibitem{ZafarRajan}
M.~Z. {Khan} and B.~S. {Rajan}, ``Single-symbol maximum likelihood decodable
  linear stbcs,'' {\em IEEE Transactions on Information Theory}, vol.~52,
  pp.~2062--2091, May 2006.

\bibitem{3tx}
B.~Ozbek, D.~Ruyet, and M.~Bellanger, ``Non-orthogonal space-time block coding
  design for 3 transmit antennas,'' {\em 19° Colloque sur le traitement du
  signal et des images, 2003 ; p. 595-598}, 01 2003.

\bibitem{frlr}
S.~S.~H. {Bidaki}, S.~{Talebi}, and M.~{Shahabinejad}, ``A full-rate
  full-diversity 2$\times$2 space-time block code with linear complexity for
  the maximum likelihood receiver,'' {\em IEEE Communications Letters},
  vol.~15, pp.~842--844, Aug 2011.

\bibitem{golden}
J.~Belfiore, G.~Rekaya, and E.~Viterbo, ``The golden code: A 2 $\times$ 2
  full-rate space-time code with non-vanishing determinants,'' {\em IEEE
  Transactions on Information Theory}, pp.~1432--1436, 2005.

\bibitem{rate2}
V.~{Vakilian} and H.~{Mehrpouyan}, ``High-rate and low-complexity space-time
  block codes for $2 \times 2$mimo systems,'' {\em IEEE Communications
  Letters}, vol.~20, pp.~1227--1230, June 2016.

\bibitem{JafarkhaniQSTBC}
H.~{Jafarkhani}, ``A quasi-orthogonal space-time block code,'' {\em IEEE
  Transactions on Communications}, vol.~49, pp.~1--4, Jan 2001.

\bibitem{WSu2004}
{W. Su}, {X. Xia}, and K.~J.~R. {Liu}, ``A systematic design of high-rate
  complex orthogonal space-time block codes,'' {\em IEEE Communications
  Letters}, vol.~8, pp.~380--382, June 2004.

\bibitem{WSu2008}
R.~{Grover}, W.~{Su}, and D.~A. {Pados}, ``An $8\times8$ quasi-orthogonal stbc
  form for transmissions over eight or four antennas,'' {\em IEEE Transactions
  on Wireless Communications}, vol.~7, pp.~4777--4785, Dec 2008.

\bibitem{Telatar99}
I.~E. Telatar, ``Capacity of multi-antenna gaussian channels,'' {\em EUROPEAN
  TRANSACTIONS ON TELECOMMUNICATIONS}, vol.~10, pp.~585--595, 1999.

\bibitem{Foschini1998}
G.~Foschini and M.~Gans, ``On limits of wireless communications in a fading
  environment when using multiple antennas,'' {\em Wireless Personal
  Communications}, vol.~6, pp.~311--335, Mar 1998.

\bibitem{EricssonMobility2018}
{Ericsson Incorporated}, ``{Ericsson Interim Mobility Report, February 2018}.''
  \url{https://www.ericsson.com/assets/local/mobility-report/documents/2018/emr-interim-feb-2018.pdf},
  2018.

\bibitem{SE2017}
K.~N. {Poudel} and S.~{Gangaju}, ``Spectral efficiency, diversity gain and
  multiplexing capacity analysis for massive mimo, 5g communications system,''
  in {\em Proc. of International Conference on Networking and Network
  Applications (NaNA)}, pp.~133--137, Oct 2017.

\bibitem{stbcfullrate2007}
S.~{Sezginer} and H.~{Sari}, ``Full-rate full-diversity 2 x 2 space-time codes
  of reduced decoder complexity,'' {\em IEEE Communications Letters}, vol.~11,
  pp.~973--975, Dec 2007.

\bibitem{Tarokh1999}
V.~{Tarokh}, H.~{Jafarkhani}, and A.~R. {Calderbank}, ``Space-time block codes
  from orthogonal designs,'' {\em IEEE Transactions on Information Theory},
  vol.~45, pp.~1456--1467, July 1999.

\bibitem{rciod}
M.~Z. {Khan}, B.~S. {Rajan}, and {M. H. Lee}, ``Rectangular co-ordinate
  interleaved orthogonal designs,'' in {\em Proc. of IEEE Global
  Telecommunications Conference (GLOBECOM)}, vol.~4, pp.~2004--2009 vol.4, Dec
  2003.

\bibitem{Tirkkonen2000}
O.~{Tirkkonen} and A.~{Hottinen}, ``Complex space-time block codes for four tx
  antennas,'' in {\em Proc. of IEEE. Global Telecommunications Conference
  (GLOBECOM)}, vol.~2, pp.~1005--1009 vol.2, Nov 2000.

\bibitem{Tirkkonen2001}
O.~{Tirkkonen} and A.~{Hottinen}, ``Improved mimo performance with
  non-orthogonal space-time block codes,'' in {\em Proc. of IEEE Global
  Telecommunications Conference (GLOBECOM)}, vol.~2, pp.~1122--1126 vol.2, Nov
  2001.

\bibitem{TarokhRank}
V.~{Tarokh}, N.~{Seshadri}, and A.~R. {Calderbank}, ``Space-time codes for high
  data rate wireless communication: performance criterion and code
  construction,'' {\em IEEE Transactions on Information Theory}, vol.~44,
  pp.~744--765, Mar 1998.

\bibitem{SuTx4}
R.~{Grover}, W.~{Su}, and D.~A. {Pados}, ``An $8\times8$ quasi-orthogonal stbc
  form for transmissions over eight or four antennas,'' {\em IEEE Transactions
  on Wireless Communications}, vol.~7, pp.~4777--4785, Dec 2008.

\end{thebibliography}
\end{document}